\documentclass[aps,prl,twocolumn,floatfix,amsmath,amssymb,showpacs,superscriptaddress,longbibliography,10pt]{revtex4-2}
\usepackage[final]{graphicx}
\usepackage[caption=false]{subfig}
\usepackage{placeins}  
\usepackage{color}
\usepackage{dcolumn} 
\usepackage{xcolor}
\usepackage{url} 
\usepackage{mathtools}
 
\usepackage{bm}
\usepackage{epsfig,psfrag,amsmath,amssymb} 
\usepackage{hyperref}
\usepackage[percent]{overpic}
\usepackage{physics}
\usepackage{booktabs} 

\hypersetup{
    colorlinks=true,
    linkcolor=blue,
    citecolor=blue,
    filecolor=magenta,      
    urlcolor=blue,
}

\setcitestyle{square}

\bibpunct{[}{]}{,}{n}{}{}

\usepackage{pdfpages} % include pdfs
\usepackage{pgffor} % for loops

\makeatletter
\AtBeginDocument{\let\LS@rot\@undefined}
\makeatother

\def\supplementfilename{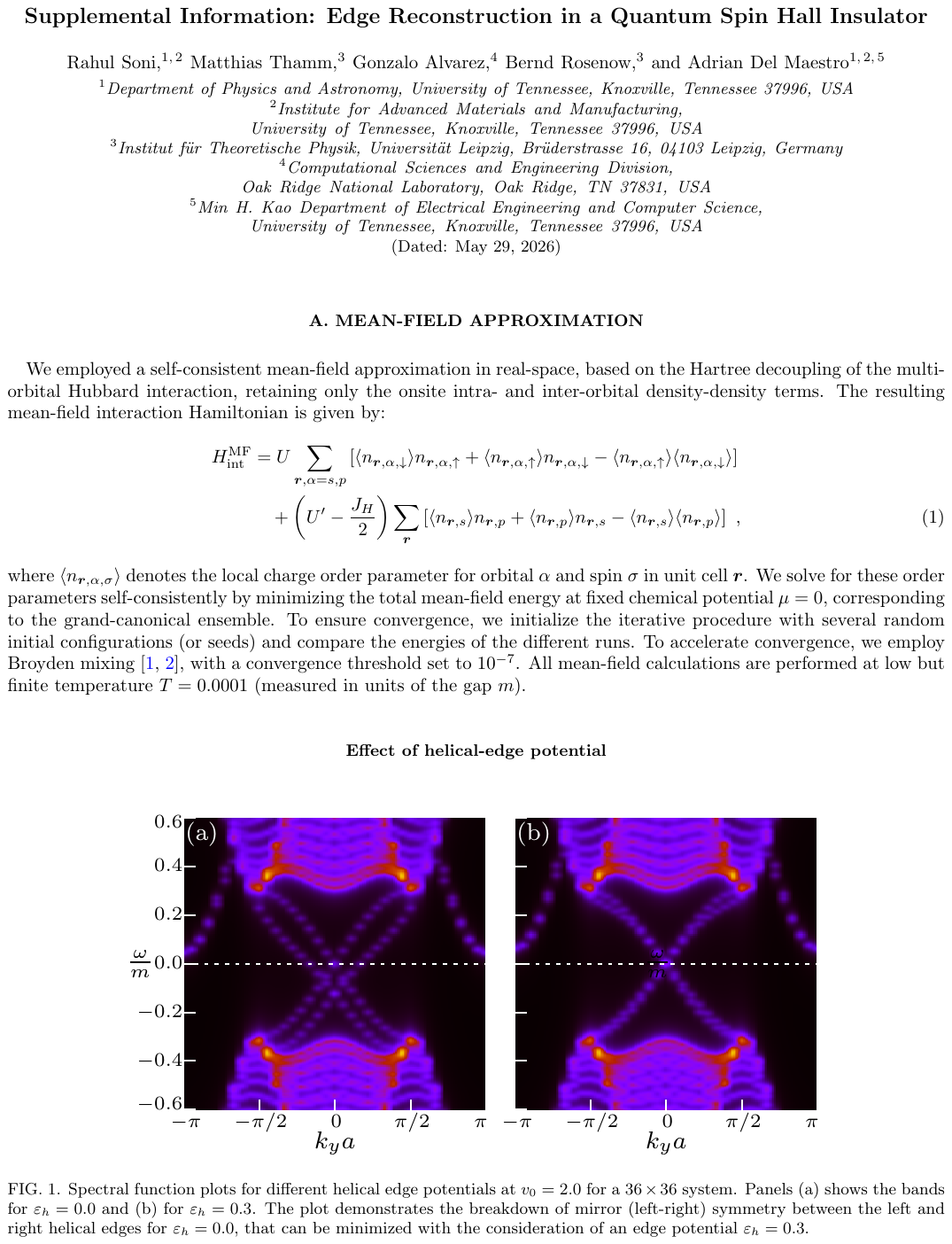}

\pdfximage{\supplementfilename}
\def\numbersupplementpages{\the\pdflastximagepages}

\newif\ifarXiv
\arXivtrue 
\begin{document}
\title{Edge Reconstruction in a Quantum Spin Hall Insulator}

\author{Rahul Soni}
\affiliation{Department of Physics and Astronomy, University of Tennessee, Knoxville, Tennessee 37996, USA}
\affiliation{Institute for Advanced Materials and Manufacturing, University of Tennessee, Knoxville, Tennessee 37996, USA}
\author{Matthias Thamm}
\affiliation{Institut f{\"u}r Theoretische Physik, Universit{\"a}t Leipzig, Br{\"u}derstrasse 16, 04103 Leipzig, Germany}
\author{Gonzalo Alvarez}
\affiliation{Computational Sciences and Engineering Division, Oak Ridge National Laboratory, Oak Ridge, TN 37831, USA}
\author{Bernd Rosenow}
\affiliation{Institut f{\"u}r Theoretische Physik, Universit{\"a}t Leipzig, Br{\"u}derstrasse 16, 04103 Leipzig, Germany}
\author{Adrian Del Maestro}
\affiliation{Department of Physics and Astronomy, University of Tennessee, Knoxville, Tennessee 37996, USA}
\affiliation{Institute for Advanced Materials and Manufacturing, University of Tennessee, Knoxville, Tennessee 37996, USA}
\affiliation{Min H. Kao Department of Electrical Engineering and Computer Science, University of Tennessee, Knoxville, Tennessee 37996, USA} 
\date{\today}

\begin{abstract}
We study interaction-driven edge reconstruction in a quantum spin Hall insulator described by the Bernevig-Hughes-Zhang model with Kanamori–Hubbard interactions using the real-space density matrix renormalization group method in both the grand-canonical and canonical ensembles.  For a two-dimensional cylinder with a smooth edge, we identify discrete particle-number transitions that lead to a spin-polarized edge state stabilized by an emergent ferromagnetic exchange interaction. The reconstruction is orbital-selective, occurring predominantly in the $s$-orbital channel. Our results reveal a microscopic mechanism for emergent fluctuating moments at the edge that could compromise the topological protection of helical edge states by time reversal symmetry. 
\end{abstract}
\maketitle

\textit{Introduction.}---The theoretical prediction of a two-dimensional quantum spin Hall (QSH) insulator protected by time-reversal symmetry was first made by Kane and Mele for graphene-like systems, introducing the notion of counter-propagating or \emph{helical} edge modes that are immune to non-magnetic backscattering \cite{kane-mele05-1,kane-mele05-2}. Building on this, Bernevig, Hughes, and Zhang (BHZ) constructed a descriptive four-band model for HgTe/CdTe quantum well heterostructures, demonstrating that an inverted band structure gives rise to gapless helical edges and a quantized conductance of $G=2e^2/h$ in the absence of magnetic fields \cite{bhz06}. Shortly thereafter, \citet{konig07} provided the first transport evidence for the quantum spin Hall effect, observing a robust $2e^2/h$ plateau at low temperatures in HgTe/(Hg,Cd)Te devices of micrometer length scales. However, subsequent experiments in longer devices have reported deviations from perfect quantization, with conductance decaying over distances as small as a few microns  \cite{Koenig.2008, Brne.2012, Koenig.2013, Hart:2014ej, Dartuaulh.2020,molenkamp19-1} and, in some cases, reporting an unexpectedly robust plateau emerging at $1e^2/h$ \cite{molenkamp19-2}. 
Beyond the original platform, QSH behavior has been observed in  InAs/GaSb heterostructures  \cite{Knez.2011, Spanton.2014, Du.2015, Nichele.2016, Strunz.2019}, monolayer WTe$_2$ \cite{Fei.2017, Wu.2018, Zhao.2020,Papaj:2024ss}, and bismuthene on SiC \cite{Reis.2017, Sthler.2019}, and Moir\'e materials \cite{Kang2024NatureFQSHMoTe2,Kang2024NanoLettDoubleQSHWSe2}.

A variety of explanations for the sensitivity of QSH conductance to experimental parameters have been put forward including the dynamical polarization of nuclear spins coupled to helical states \cite{Lunde.2012, DelMaestro.2013, Hsu.2017, Russo.2018, Hsu.2018}, incoherent electromagnetic noise \cite{Alicea:2018bs}, the presence of Rashba spin-orbit coupling \cite{strom2010edge,Geissler.2014, Xie.2016, Kharitonov.2017,Strunz.2019,Mondal.2022} combined with interaction mediated effects \cite{Bairam.2019, Chou.2018}, scattering from magnetic impurities \cite{Maciejko:2009ke, Kimme.2016, Kurilovich.20171, Kurilovich.20172, Altshuler.2013, Vezvaee.2018,Zheng.2018, Pashinsky.2020,Vannucci:2020cq,Dietl23}, and scattering from  non-magnetic impurities or electron puddles \cite{Lezmy.2012,Varynen.2013,Gefen.2014, Novelli.2019}.

A more recent scenario to explain transport anomalies is the reconstruction of the charge density near the boundary itself, i.e., additional localized low-energy modes or extra one-dimensional edge channels may emerge proximate to the helical edge due to electrostatics, confinement, or interactions.  
This physics has been extensively studied in integer and fractional quantum Hall systems \cite{halperin93,wan02,wan03,yang03,Gefen22,Lotric:2025er,rao_and_murty21,rao_gefen17}. Prior theoretical work on edge reconstruction in QSH systems, largely within mean-field frameworks \cite{gefen17,rosenow22,trauzettel17,hyart23}, has demonstrated that once multiple channels exist, time reversal symmetry only protects each Kramers pair individually but does not forbid scattering between different pairs. Moreover, enhanced edge density of states can even trigger local moment formation (ferromagnetism), breaking time reversal symmetry outright. Taken together, these effects provide backscattering mechanisms absent in the single-channel case, giving rise to a mesoscopic  length scale that limits ballistic transport. However, a fully interacting real-space understanding of the effects of edge reconstruction is still lacking. Moreover, while the density matrix renormalization group (DMRG) method \cite{white92,schollwock05} has recently been established as a powerful tool for characterizing topological phases in idealized BHZ models \cite{soni24}, its application to mesoscopic systems with reconstructed edges remains largely unexplored.

In this work, we move beyond the limitations of previous mean-field and effective model approaches by applying DMRG to a fully interacting, multi-orbital BHZ  model. This constitutes a controlled microscopic treatment of realistic mesoscopic boundaries that preserves global time-reversal symmetry. Our exact many-body analysis reveals an orbital-selective edge reconstruction and isolates an emergent ferromagnetic exchange interaction that stabilizes magnetic moments on the edge. This is in stark contrast to mean-field approaches that are biased towards magnetic symmetry breaking states on the edge even in the absence of edge reconstruction. 
By providing a fully interacting, real-space mechanism for the generation of fluctuating magnetic moments at the boundary, our results establish a  microscopic link between the topological properties of the bulk and the mesoscopic backscattering mechanisms that compromise ballistic transport in experimental devices~\cite{Koenig.2008,Brne.2012,Koenig.2013,Strunz.2019}. 

\begin{figure}[!t]
\includegraphics[width=\linewidth]{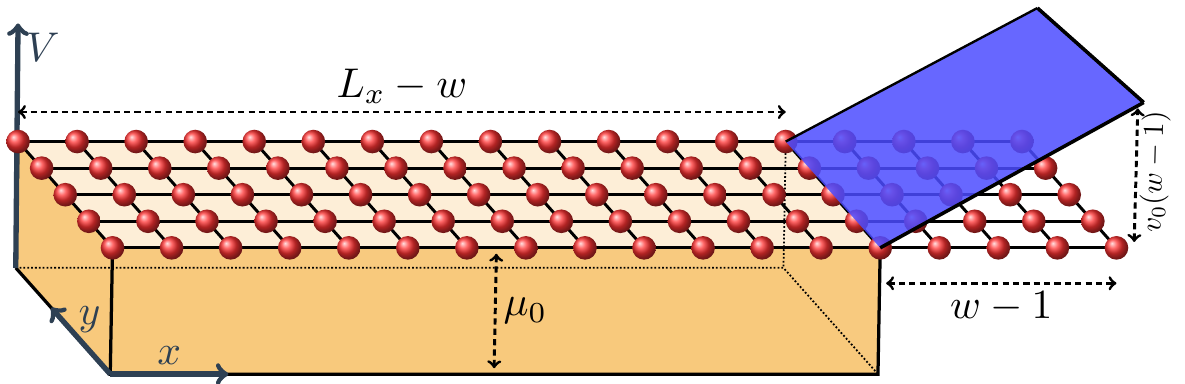}
\caption{Potential profile illustrating the combined ionic and confining potentials on a 2D lattice. The confining potential linearly increases in the extended edge region over a width $w$ with slope $v_0$ from the left helical edge at site $r_x = L_x-w$. The ionic potential (orange) is uniformly set as the negative of the atomic half-filling chemical potential $\mu_0$. We take periodic boundary conditions in the $y$-direction.} \label{Fig: lattice pot profile}
\end{figure}

\textit{Model}---We study an interacting real-space BHZ model on a two-dimensional cylinder with periodic boundary conditions along $y$ and mixed boundaries along $x$: open on the left edge and with smooth potential confinement on the right.  The Hamiltonian is given by:
\begin{equation}
H = H_{\vphantom{i^i}0} + H_{\vphantom{i^i}\rm int} + \sum_{\vb*{r}} V(r_x) n_{\vb*{r}}
\end{equation}
where $H_{\vphantom{i^i}0}$ is the real-space non-interacting BHZ model, $H_{\vphantom{i^i}\rm int}$ is the multi-orbital Kanamori-Hubbard interaction \cite{kanamori63,soni22}, and $V(r_x)$ captures the geometric details of the boundary with $n_{\vb*{r}}$ the total electron density at site $\vb*{r}$, including orbital and spin degrees of freedom. This explicitly breaks the mirror (left-right) symmetry but does not impact the global time-reversal symmetry. 

The non-interacting Hamiltonian is \cite{soni24,trauzettel13,ohta13,tkng19}:
\begin{align}
H_{\vphantom{i^i}0} &= -m \sum_{\vb*{r},\alpha} n_{\vphantom{i^i}\vb*{r},\alpha} (\tau^z)_{\vphantom{i^i}\alpha\alpha} + B \sum_{\substack{\vb*{r},\alpha,\sigma\\ i=x,y}} c^{\dagger}_{\vb*{r},\alpha,\sigma}(\tau^z)_{\vphantom{i^i}\alpha\alpha}c_{\vphantom{i^i}\vb*{r}+\hat{\vb*{e}}_{i},\alpha,\sigma}  \nonumber \\
& + \frac{A}{2}\sum_{\substack{\vb*{r},\sigma\\ i = x,y }}\sum_{\substack{\alpha,\beta\\ \alpha\neq\beta}} c^{\dagger}_{\vb*{r},\alpha,\sigma}(D_\sigma ^i)_{\vphantom{i^i}\alpha\beta}c_{\vphantom{i^i}\vb*{r}+\hat{\vb*{e}}_{i},\beta,\sigma}  + \mathrm{h.c.} \ ,\label{Eqn: Non-int. Ham}
\end{align}
where the first term describes the onsite energy with $0<\lvert \frac{m}{2B}\rvert<2$ corresponding to the topological phase and $\lvert \frac{m}{2B}\rvert>2$ the trivial phase~\cite{trauzettel13}. The second term is the nearest neighbor hopping, and the third captures spin-orbit coupling with mixing matrices $(D^{x}_{\sigma}) = (-1)^{\sigma}(\mathfrak{i}\tau^{x})$ and $(D^y_{\sigma})= (-\mathfrak{i}\tau^{y})$. Here, $\vb*{r}=(r_{x},r_{y})$ represents the unit cell vector with components $r_{x}$ and $r_{y}$ along the lattice vectors $a\hat{\vb*{e}}_x$ and $a\hat{\vb*{e}}_y$, respectively, $\alpha,\beta=s,p$ denotes the orbital indices, and $\sigma= \uparrow, \downarrow$ represents the $z$-axis spin projection of an electron.  The $\tau^{i}$ are Pauli matrices in orbital space. The factor $(-1)^{\sigma}=-1(1)$ for $\sigma=\uparrow(\downarrow)$, respectively.

The multi-orbital Hubbard interaction is \cite{soni24,kanamori63,soni22}:
\begin{align}
H_{\vphantom{i^i}\rm int} &=  U \sum_{\vb*{r},\alpha} n_{\vphantom{i^i}\vb*{r},\alpha,\uparrow}n_{\vphantom{i^i}\vb*{r},\alpha,\downarrow} + \left(U'-\frac{J_{H}}{2}\right)\sum_{\vb*{r}}n_{\vphantom{i^i}\vb*{r},s}n_{\vphantom{i^i}\vb*{r},p} \nonumber \\
&  - 2J_{H}\!\sum_{\vb*{r}} \vb*{S}_{\vphantom{i^i}\vb*{r},s}\!\cdot\!\vb*{S}_{\vphantom{i^i}\vb*{r},p} +J_{H}\sum_{\vb*{r}}\left( P_{\vb*{r},s}^{\dagger}P^{\phantom{\dagger}}_{\vb*{r},p} + \mathrm{h.c} \right),\label{Eqn: MOHI}
\end{align}
where $U'=U-2J_{H}$, $\vb*{S}_{\vb*{r},\alpha}$ is the spin operator, and $P_{\vb*{r},\alpha}=c_{\vb*{r},\alpha,\uparrow}c_{\vb*{r},\alpha,\downarrow}$ the pair-annihilation operator.  
Throughout this paper, we measure lengths in units of the lattice spacing $a$ %set the lattice spacing to $a=1$ 
and use the gap size $m$ as the unit of energy, working in a fixed parameter regime: $A=0.3, B=0.5, U=1.0$ and $J_H=0.25$. This corresponds to the paramagnetic topological insulator phase \cite{soni24}.

\begin{figure}[t]
\hspace*{-0.3cm}\includegraphics[width=\linewidth]{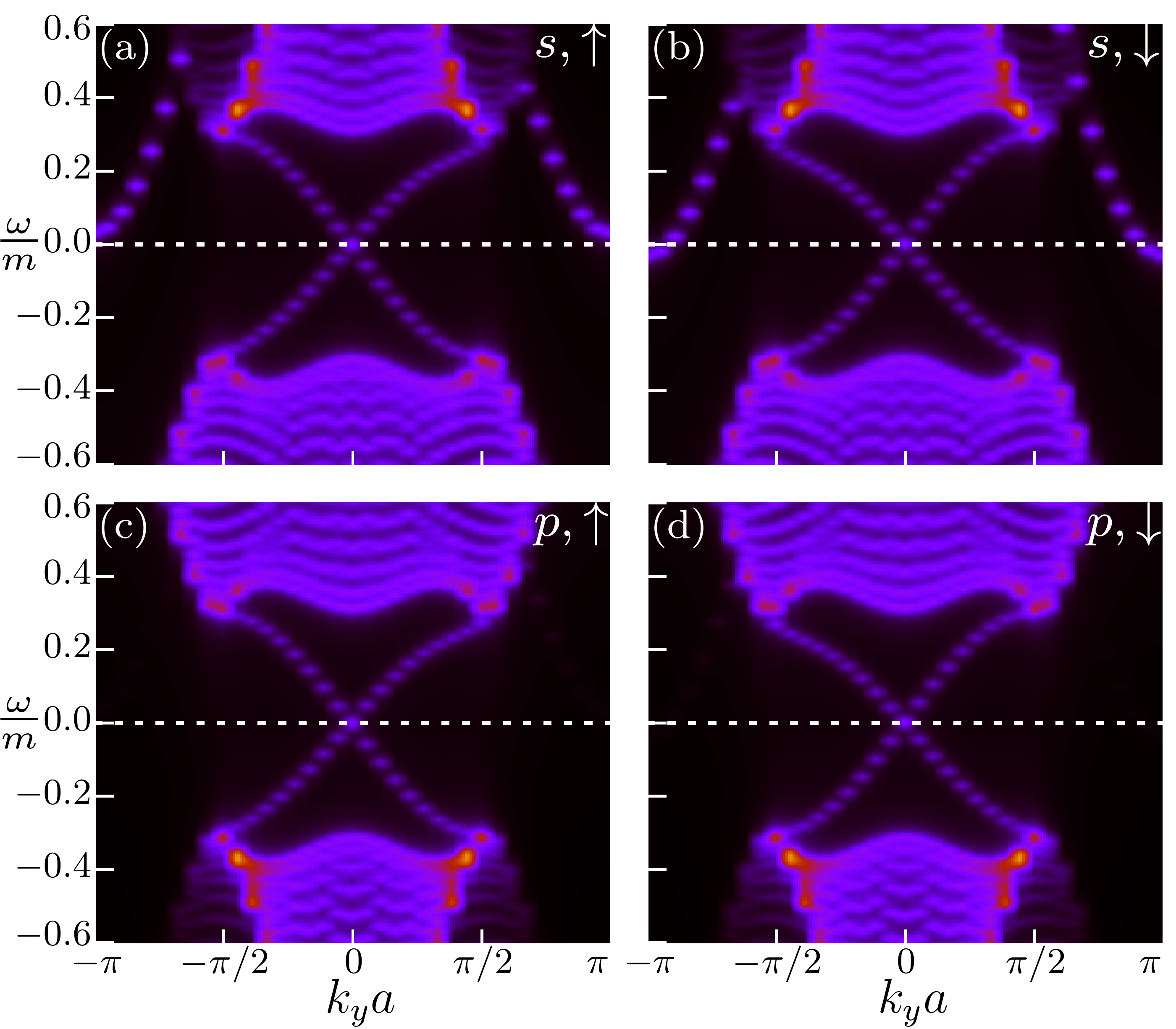}
\caption{Orbital and spin resolved single-particle spectral function from self-consistent mean-field calculations on a 36$\times$36 cylinder with confining width $w=6$, confining potential slope $v_0=1.90$, and $\varepsilon_h = 0.3$. Panels (a) and (b) show the energy bands associated with the orbital-$s$ channel for spin-$\uparrow$ and spin-$\downarrow$ particles, respectively. Panels (c) and (d) display the corresponding bands for the orbital-$p$ channel. The plot demonstrates the orbital-selective nature of edge reconstruction.} \label{Fig: MF spectral OS resolved}
\end{figure}

To model a smooth boundary at one edge, as depicted in Fig.~\ref{Fig: lattice pot profile}, we use an electrostatic potential $V(r_x)$ that increases linearly from the right helical edge over a width $w$, combined with an ionic potential at positions up to and including the right helical edge:
\begin{equation*}
V(r_x)
  = \begin{cases}
-\mu_{0} + \varepsilon_h \, \delta_{r_x,L_x - w} \ & 0\leq r_x \leq L_x-w \\
v_{0} (r_x - L_x +w) & L_x-w< r_x\leq L_x-1
\end{cases} \label{Eqn: Conf-ionic pot}
\end{equation*}
where \(\mu_0=(3U-5J_H)/2\) ensures charge neutrality in the bulk, $v_0$ is the slope of the confinement potential, and $L_x$ is the total number of lattice sites along the cylinder.  The additional additive term $\varepsilon_h$, acting at the right helical edge, prevents the accumulation of extra charge due to a softening of the confinement potential. In all our calculations, the value of $\varepsilon_h$ is chosen such that the left and right helical edges are similar (e.g.\@ no splitting of the Dirac point in Fig.~\ref{Fig: MF spectral OS resolved}, same particle number, see supplement for more details \cite{supplement}).

\textit{Mean-field results.}---To complement our DMRG study, we performed mean-field calculations on two-dimensional cylinders, details are provided in the supplemental material \cite{supplement}. We evaluate the orbital and spin-resolved spectral function for momentum $k_y$ around the cylinder. Our results in Fig.~\ref{Fig: MF spectral OS resolved} reveal that at the onset of edge reconstruction for $v_0 = 1.9$, signaled by the appearance of a fermi pocket at the Brillouin zone edge, a spin splitting emerges exclusively in the $s$-orbital bands, while the $p$-orbital bands remain mostly unaffected.  Although spin splitting associated with edge reconstruction has been previously reported \cite{gefen17,rosenow22}, its orbital character, specifically its confinement to the $s$-channel, was missed due to a lack of orbital information. This orbital selectivity follows directly from the onsite energy term in Eq.~\eqref{Eqn: Non-int. Ham}, which assigns onsite energies $-m$ and $+m$ to $s$-orbital and $p$-orbital occupations, respectively, such that one can obtain reconstruction for orbital-$p$ by changing the crystal-field term $m\rightarrow -m$, see supplement~\cite{supplement}.

\begin{figure}[t]
\centering
\includegraphics[width=\linewidth]{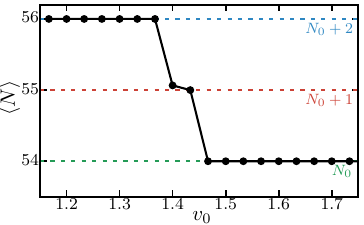}
\caption{Total number of particles $\expval{N}$ as a function of the slope of confining potential $v_0$ for a $12\times 3$ cylinder with confining width $w=4$, computed using grand-canonical DMRG. 
At large values of $v_0$, the particle number stays on a plateau where the bulk is at half-filling.   Reduction of $v_0$ leads to particle accumulation in the confined region, signaling the onset of edge reconstruction.} 
\label{Fig: N vs V0 GCE}
\end{figure}

\begin{figure}[!t]
\centering
\includegraphics[width=\linewidth]{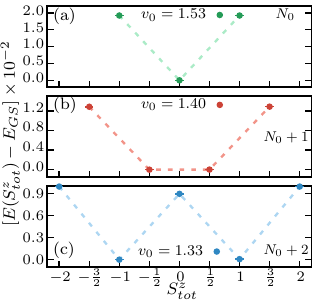}
\caption{Canonical DMRG ground-state energies as a function of total spin $S^z_{\rm tot}$ on a $12\times 3$ cylinder with confining width $w=4$. Panels (a), (b), and (c) corresponds to the $N_0$, $N_0+1$, and $N_0+2$ particle sectors, with confining potentials $v_0=1.53$, $1.40$, and $1.33$, respectively. The ground state lies in the $S^z_{\rm tot}=0$, $\pm \tfrac{1}{2}$, and $\pm 1$ sectors for $N_0$, $N_0+1$ and $N_0+2$. In the $N_0+2$ case, the system favors a degenerate spin-polarized configuration, indicating an energetic preference for ferromagnetic alignment of spins at the reconstructed edge.} \label{Fig: GSE vs Sz CE}
\end{figure}

\textit{DMRG results}--We focus on quasi one-dimensional cylinders of size $L_x\times 3$, working in both the canonical and grand-canonical ensembles to investigate the microscopic nature of edge reconstruction. We first perform grand-canonical DMRG calculations \cite{ITensor,ITensor-r0.3} to identify the edge reconstruction regime by investigating changes in the particle number as the slope of the confining potential, $v_0$, is decreased. Once the value of $v_0$ for which additional particles are energetically favorable on the reconstructed edge is known, we carried out high-precision canonical DMRG calculation \cite{alvarez0209,alvarez0310,alvarez08} across adjacent particle sectors to determine the global ground states and analyze the spin and orbital character in real-space. We considered up to $2000$ DMRG states in the grand-canonical and up to $13000$ states in the canonical ensemble calculations, imposing a maximum truncation error of $10^{-5}$. The results presented here correspond to $12\times 3$ cylinders with confining width $w=4$ and $\varepsilon_h=0.4$. This system size is sufficient to decouple the edges while maintaining a well-defined bulk. Finite size effects are explored in the supplement \cite{supplement}. 

Figure~\ref{Fig: N vs V0 GCE} describes the change in the average total number of particles $\expval{N} = \expval{\sum_{\vb{r}}n_{\vb{r}}}$ as a function of the slope of the confining potential $v_0$, computed within the grand-canonical ensemble. As $v_0$ decreases from infinity, the system transitions from the hard-wall regime to the reconstructed regime. Initially, the system is stable at half-filling forming a broad plateau in $\expval{N} = N_0$, which for our $12\times 3$ system with confining width $w=4$ corresponds to $N_0=(9\cdot 3\cdot 2 \cdot 2) / 2 =54$ particles (2 spins, 2 orbitals per site).  As $v_0$ is reduced, we observe a sharp jump in particle number, signaling the onset of edge reconstruction, with the additional charge now localized in the confined region (as demonstrated below in Fig.~\ref{Fig: del ni vs rx CE}). This first jump corresponds to the $N_0+1$ particle sector. Further reduction of the confining potential induces another discrete jump in $\expval{N}$, yielding the $N_0+2$ sector, where two particles predominantly accumulate in the reconstructed region. 

Having identified three distinct particle number plateaus associated with edge reconstruction, we next turn to a canonical DMRG analysis to determine ground state properties for $N_0$, $N_0+1$, and $N_0+2$ corresponding to $v_0=1.53$, $1.40$, and $1.33$ respectively. In Fig~\ref{Fig: GSE vs Sz CE} we show the ground state energy for different values of the fixed spin quantum number $S^z_{\rm tot}$. For $N_0$ particles, the minimum occurs at $S^z_{\rm tot}=0$, expected for a half-filling. For $N_0+1$ particles, the lowest energy states are doubly-degenerate between the $S^z_{\rm tot}=\pm\frac{1}{2}$ sectors, reflecting the presence of an unpaired spin localized at the reconstructed edge. Surprisingly, for $N_0+2$ electrons, the ground states lie in the $S^z_{\rm tot}=\pm 1$ sectors, with the $S^z_{\rm tot}=0$ state appearing at higher energy, indicating a spin polarized edge with non-zero magnetic moment. This energetic preference supports an emergent ferromagnetic alignment of the two added edge-localized particles, consistent with an exchange-driven edge reconstruction as described below.

\begin{figure}[!t]
\centering
\includegraphics[width=\linewidth]{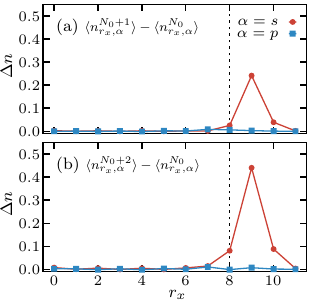}
\caption{Real-space charge density differences depicting how the reconstructed states are localized in the confinement region. Panel (a) shows the difference between the density profiles of the $N_0+1$ and $N_0$ sectors; (b) shows the difference between $N_0+2$ and $N_0$ sectors. We average over the degenerate ground states for the cases of $N_0+1$ and $N_0+2$ electrons and along the periodic $y$-direction.  The vertical dashed lines depict the location of the right-helical edge.} \label{Fig: del ni vs rx CE}
\end{figure}

To gain further insight into the spatial and orbital character of these added electrons, we analyze the resulting real-space charge and magnetic properties of the helical and reconstructed states. Figure~\ref{Fig: del ni vs rx CE} shows the expectation value of  the spatial distribution of added charge through the difference in orbital resolved $y$-averaged charge densities, $\expval{n_{r_x,\alpha}}=\frac{1}{L_y}\sum_{r_y} \expval{n_{\vb*{r},\alpha}}$, between the ground-states of different particle number sectors. 
Panel (a) confirms that when adding a single electron to the system, it is localized in the confinement region $r_x > L_x-w$. With the addition of a second electron, the $N_0+2$ ground state now demonstrates some leakage of density onto the helical edge.  
In both cases, the excess density appears almost entirely in the orbital-$s$ channel, with negligible contributions to the orbital-$p$ channel, confirming the orbital-selective nature of edge reconstruction, consistent with our mean-field spectral analysis in Fig.~\ref{Fig: MF spectral OS resolved}.

\begin{figure}[!t] 
\centering
\includegraphics[width=\linewidth]{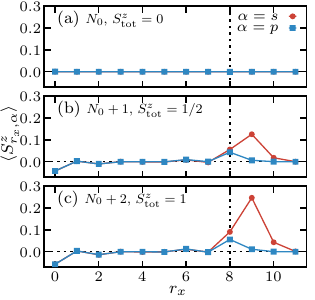}
\caption{Real-space orbital and site-resolved magnetization for the ground states of different particle number sectors with fixed $S^z_{\rm tot}$ quantum numbers. Panel (a) shows magnetization for $N_0$ particles in the $S^z_{\rm tot}=0$ state; panel (b) for $N_0+1$ in the $S^z_{\rm tot}=+\frac{1}{2}$ state; and panel (c) for $N_0+2$ in the $S^z_{\rm tot}=+1$ state. The vertical dashed line depicts the right-helical edge. We observe a  $p$-channel magnetic response at the location of the helical edge, indicating that there may be an interaction coupling between reconstructed and helical edge. \label{Fig:_szi_vs_rx_CE}} 
\end{figure} 

The magnetic properties of the edge are shown in Fig.~\ref{Fig:_szi_vs_rx_CE}, which depicts the orbital and site resolved $y$-averaged magnetization $\langle S^z_{r_x,\alpha}\rangle$ for the ground states of the three particle sectors of interest. For $N_0$ particles (half-filing) the total and local magnetization are zero as seen in panel (a).  
As the confinement potential is softened and the system transition to the $N_0+1$ state in panel (b), the additional electron causes a local moment in the confinement region mostly in the $s$-channel.  However, we also observe an interaction induced increase in the $p$-channel moment directly on the right helical edge with a concomitant response on the left helical edge due to the restriction to $S^z_{\rm tot}=+\frac{1}{2}$. Panel (c) for $N_0+2$ and $S^z_{\rm tot}=+1$ shows a further enhancement of correlated local moments in the confinement region and helical edges, indicative of a coupling between magnetic moments on the reconstructed and on the helical edge. Thus, the observed ferromagnetic alignment of the two added edge-localized electrons provides real-space evidence of an exchange driven edge magnetism associated with edge reconstruction.

\emph{Emergent Spin Hamiltonian}---DMRG calculations supplemented by the exact diagonalization (ED) analysis for two particles on a single rung (see end matter) allow us to infer an effective exchange Hamiltonian between the two spin-$\frac{1}{2}$ particles localized in the reconstructed region for $N_0+2$. From ED, we find a singlet-triplet splitting $\Delta E=0.2167$ which is considerably larger than the spin exchange anisotropy $\Delta E_a=0.0089$ (due to the combination of Hund's coupling and spin dependent hopping, shown in  Fig.~\ref{Fig: GSE vs Sz CE}). Taken together this physics can be described by $H_{\rm exc} = -J\left(\mathbf{S}_i^{\phantom z} \cdot \mathbf{S}_j^{\phantom z}\right) - \Delta S^z_i S^z_j$
where the first term describes the magnitude of a singlet-triplet splitting, and the second term describes the anisotropy between the components of the triplet state with $S^z=0$ and $S^z=\pm 1$. This simple picture yields $J= \Delta E \simeq 0.2167$ and for our $12 \times 3$ cylinder, $\Delta =2\Delta E_a \simeq 0.0178$, indicating a finite anisotropic ferromagnetic coupling that stabilizes a spin-polarized edge configuration. For a finite size scaling analysis of these couplings see the supplemental material \cite{supplement}.

\textit{Discussion}---Our results identify a microscopic route to edge magnetism in a quantum spin Hall system: sample boundary softening through a spin-independent electrostatic potential drives an interaction-induced, orbital-selective spectral reconstruction in which the $s$-channel predominantly admits charges and exhibits spin splitting, while the $p$-channel remains mostly incompressible. The orbital selectivity is tied to the crystal-field splitting and thus a $p$-orbital polarized edge reconstruction could be realized via strain tuning. This mechanism is tied to the topological band inversion and can be readily applied to systems with $p$-$d$ \cite{Qian2014,Chadov2010} or $d$-$f$ ~\cite{Neupaneetal2013,Hung2020} orbital character.

Plateaus followed by jumps in the total particle number in the grand canonical ensemble signal generic edge reconstruction with canonical calculations at $N_0$, $N_0+1$, $N_0+2$ linking the charge steps to real-space spin densities.  The latter can be used to diagnose the presence of a finite ferromagnetic exchange coupling from the energy differences between different total spin sectors, explaining the observed ground-state spin polarization at the reconstructed edge even though the Hamiltonian preserves time-reversal symmetry.  
While we study a finite lattice model, from  a Luttinger–liquid viewpoint a ferromagnetic tendency can arise when the spin stiffness (the coefficient of the quadratic gradient term) changes sign, with a higher-gradient term stabilizing the field theory and yielding unconventional dynamical scaling \cite{Yang:2004ft}. However, the Lieb–Mattis theorem \cite{Mattis:1962tf} forbids true ferromagnetism for a single occupied sub-band with interactions shorter than the interparticle spacing, and in the opposite limit of long-range Coulomb forces a ferromagnetic state competes with Wigner crystallization \cite{Schulz:1993wc}. We therefore expect that the mesocopic devices in the current era of experiments should exhibit fluctuating edge moments rather than stable long-range ferromagnetism with implications for the protection of ballistic transport.

\textit{Code and Data Availability}---We employ \texttt{ITensor} \cite{ITensor,ITensor-r0.3} and \texttt{DMRG++} \cite{alvarez08}  for DMRG simulations. All data, code, and analysis scripts that support the findings of this study are available online \cite{paperrepo}.

\textit{Acknowledgements}---R.~S., and A.~D. acknowledge support from the U.~S. Department of Energy, Office of Science, Office of Basic Energy Sciences, under Award No. DE-SC0022311. B.~R. acknowledges support from the German Research Foundation under grant RO 2247/16-1 and the hospitality of the University of Tennessee, where a portion of this work was performed. G.~A. acknowledges support from the U.~S. Department of Energy, Office of Science, National Quantum Information Science Research Centers, Quantum Science Center.  

\FloatBarrier
\bibliographystyle{apsrev4-1} 
%\bibliography{Reference_EdgeRecon_BHZ}

\newpage
\onecolumngrid \vspace{0.5cm}
 \begin{center}{\large\textbf{End Matter}}\end{center} \vspace{0.1cm} 
\twocolumngrid \noindent

\emph{Exact diagonalization of a three site ring}---Both our mean-field and DMRG calculations indicate that a spin splitting emerges when more than one particle occupies the confined region ($N_0+2$), with the added charge predominantly localized near the right-helical edge. Specifically, in the case of two additional particles, they become localized at the rung adjacent to the right-helical edge, where the confining potential is minimal (but non-zero). To isolate the origin of this mechanism and quantify the scale of the singlet-triplet gap, we consider an exactly solvable system in which two particles are confined entirely to a three site ring with two orbitals per site.  The reduced Hamiltonian along the $y$ direction is $H_{\rm ring}=H_s+H_p+H_{sp}$ where
\begin{align}
H_{\alpha} &= (\tau^z)_{\alpha\alpha} B\sum_{r_y,\sigma}\left(c^{\dagger}_{r_y,\alpha,\sigma}c_{r_y+1,\alpha,\sigma} + \mathrm{h.c.}\right) \nonumber \\
& \quad -  (\tau^z)_{\alpha\alpha} m \sum_{r_y} n_{r_y,\alpha} + U\sum_{r_y}n_{r_y,\alpha,\uparrow}n_{r_y,\alpha,\downarrow}
\label{Eqn. Reduced aa Hamiltonian} \\
H_{sp} &= \frac{A}{2}\!\sum_{r_y,\sigma}\left(c^{\dagger}_{r_y,s,\sigma}c_{r_y+1,p,\sigma} - c^{\dagger}_{r_y,p,\sigma}c_{r_y+1,s,\sigma} + \text{h.c.} \right) \nonumber \\
& + \qty(U'-\tfrac{J_H}{2})\!\!\sum_{r_y}n_{r_y,s}n_{r_y,p} -2 J_H\!\!\sum_{r_y}\vb*{S}_{r_y,s}\cdot \vb*{S}_{r_y,p}\nonumber \\
& + J_H \sum_{r_y}\left(P^\dagger_{r_y,s}P_{r_y,p}^{\phantom \dagger} + \mathrm{h.c.}\right).
\label{Eqn. Reduced sp Hamiltonian}
\end{align}
We have omitted the confinement potential $v_0\sum_{r_y,\alpha,\sigma}n_{r_y,\alpha,\sigma}$ acts as a uniform chemical potential that can be absorbed as an energy offset. 
\begin{table}[t]
    \centering 
    \caption{Energy eigenvalues for the lowest six eigenstates of the three site ring and their total spin quantum number ($S_{\rm tot}$). The second column uses the parameters from the main text. In the third column, we set the interaction to zero ($U=U'=J_H=0$). We find that in the non-interacting case the ground state is six-fold degenerate, containing both singlet and triplet states at the same energy. Interactions lift this degeneracy and stabilize a spin triplet ground-state. We measure energies in units of the gap $m$ and lengths in units of the lattice spacing $a$.}
    \begin{tabular}{c @{\hspace{1.1cm}} c @{\hspace{1.1cm}} c @{\hspace{0.2cm}}} 
        \toprule
        $S_{\rm tot}$  & energy & energy, \\ 
        & &  $U=J_H=0$\\
        \midrule
       $1$ &  $-3.04242$  & $-3.04467$ \\ 
       $1$  &  $-3.04242$  & $-3.04467$ \\
       $1$  &  $-3.04242$  & $-3.04467$ \\
       $0$  &  $-2.82576$  & $-3.04467$ \\
       $0$  &  $-2.82576$  & $-3.04467$ \\
       $0$  &  $-2.47194$  & $-3.04467$ \\ 
        \bottomrule
    \end{tabular}
    \label{tab:endmatter_effectivemodelEDgroundstates}
\end{table} 

The resulting Hilbert space has dimension $\abs{H_{\rm ring}} = \binom{12}{2}=66$, and we perform exact diagonalization for the same model parameters as in the main text.  We show the low-lying spectrum in Table~\ref{tab:endmatter_effectivemodelEDgroundstates} resolved by total spin quantum number ($S_{\rm tot}$). We find a three fold degenerate spin triplet as ground state for the fully interacting ring geometry separated by a finite gap with the singlet. In the absence of interaction ($U=U'=J_H=0$), the ground-state becomes six fold degenerate due to the degeneracy of the lowest single-particle momentum states, resulting in both singlet and triplet at the same energy. The interaction therefore lifts this degeneracy and stabilizes a triplet ground state, depicting the vital role of interactions in the formation of magnetic moments at the reconstructed edge.

When the spatial orientation of the ring is changed from $y$ to $x$, the orbital-mixing structure ($A\ne 0$) of the non-interacting BHZ Hamiltonian breaks SU(2) symmetry. While the triplet ground state persists for two interacting particles, the triplet manifold becomes internally split, with the Hund's coupling favoring $S^z=\pm1$ over $S^z=0$ as the ground state, thereby selecting an easy-axis direction for the total spin. The corresponding result for a three-site ring oriented along $x$ with two particles is shown in Table.~\ref{tab:supplement_effectivemodelEDgroundstates}. Thus, the interacting BHZ model at fractional filling exhibits fluctuating moments with anisotropic spin exchange.
\begin{table}[t]
    \centering 
    \caption{Energy eigenvalues for the lowest six eigenstates of the three site ring with two particles oriented along $x$ direction corresponding to their total spin-z quantum number ($S^z_{\rm tot}$). The second column uses the parameters from the main text. In the third and fourth column, we set $J_H=0$ and $A=0$, respectively. We find that there is a third degenerate ground state with $S^z_{\rm tot}=0$ if either the Hund's coupling $J_H$ or the orbital-mixing $A$ are set to zero.}
    \begin{tabular}{c @{\hspace{0.8cm}} c @{\hspace{0.8cm}} c @{\hspace{0.8cm}} c @{\hspace{0.2cm}}} 
        \toprule
        $S^z_{\rm tot}$  & energy & energy, $J_H=0$ & energy, $A=0$ \\
        \midrule
       $-1$ &  $-3.04242$  & $-3.03734$ & $-3.0\phantom{0000}$\\ 
       $\phantom{-}1$  &  $-3.04242$  & $-3.03734$ & $-3.0\phantom{0000}$\\
       $\phantom{-}0$  &  $-3.03881$  & $-3.03734$ & $-3.0\phantom{0000}$\\
       $\phantom{-}0$  &  $-2.82791$  & $-2.82259$ & $-2.78228$\\
       $\phantom{-}0$  &  $-2.82791$  & $-2.82259$ & $-2.78228$\\
       $\phantom{-}0$  &  $-2.47194$  & $-2.46787$ & $-2.42069$\\ 
        \bottomrule
    \end{tabular}
    \label{tab:supplement_effectivemodelEDgroundstates}
\end{table} 

\emph{Emergent Spin Hamiltonian}---The singlet-triplet splitting $\Delta E=0.2167$ in Table~\ref{tab:endmatter_effectivemodelEDgroundstates} is considerably larger than the spin exchange anisotropy that we obtained from our canonical DMRG calculations $\Delta E_a=0.0089$ (Fig.~\ref{Fig: GSE vs Sz CE}) for two additional particles in the reconstructed region. 

Based on this combined ED and DMRG analysis, we suggest an effective emergent exchange Hamiltonian for the $N=2$ reconstructed edge with form:
\begin{align}
    H_{\rm exc} & = -J\left(\bm{S}_i^{\phantom z} \cdot \bm{S}_j^{\phantom z}\right) - \Delta S^z_i S^z_j\, .\label{Eqn: Effective exchange}
\end{align}
The first term describes the origin of the observed singlet-triplet splitting, whereas the second term describes the anisotropy between the $S^z=0$ and $S^z=\pm 1$ states. For two particles, within this simple picture we obtain the following eigenvalues: $E_{S^z=\pm 1}=\frac{-J-\Delta}{4}$, $E_{S^z=0,{\rm triplet}} = \frac{-J+\Delta}{4}$, and $E_{S^z=0,{\rm singlet}} = \frac{3J+\Delta}{4}$. From our DMRG analysis we obtain $\Delta E_a = E_{S^z=0,{\rm triplet}} - E_{S^z=\pm 1}$ which allows us to estimate $\Delta = 2\Delta E_a \simeq 0.0178$. The ED singlet-triplet splitting is described by the limit $\Delta \approx 0$ which gives $J = \Delta E \simeq 0.2167$.
\ifarXiv
\clearpage 
\onecolumngrid
    \foreach \x in {1,...,\numbersupplementpages}
    {   
        \includepdf[pages={\x}]{\supplementfilename}
    }
\fi 
\end{document}